\documentclass[pra,twocolumn,groupedaddress,noshowpacs,checklength,amssymb,amsmath]{revtex4}

\usepackage{bm}
\usepackage{amsmath}
\usepackage{amssymb}
\usepackage{amsthm}
\usepackage{amsfonts}
\usepackage{dsfont}
\usepackage{graphicx}
\usepackage{color}
\usepackage[colorlinks=true,urlcolor=blue,linkcolor=green,citecolor=blue]{hyperref}

\begin{document}


\title{Equilibrium Charge Density on a Thin Curved Wire}


\author{M. Hossein Partovi}
\email[Electronic address:\,\,]{hpartovi@csus.edu} \affiliation{Department of Physics and Astronomy, California State
University, Sacramento, California 95819-6041}
\author{David J. Griffiths}
\email[Electronic address:\,\,]{griffith@reed.edu} \affiliation{Physics Department, Reed College, Portland, Oregon 97202}



\begin{abstract}
Equilibrium charge distribution on a thin, straight conductor has been actively discussed in recent years, starting with
Griffiths and Li \cite{djgyli}, who suspected a uniform distribution, and culminating with Maxwell's insightful words \cite{jcm} as
reported by Jackson \cite{jdj2}.  The present work addresses the electrostatic problem of a thin, curved, cylindrical conductor,
or a conducting \textit{filament}, and shows that the corresponding linear charge density slowly tends to uniformity as the
inverse of the logarithm of a characteristic parameter which is the ratio of the diameter to the smaller of the length and
minimum radius of curvature of the filament.  An alternative derivation of this result directly based on energy minimization is
developed.  These results are based on a general asymptotic analysis of the electric field components and potential near a
charge filament in the limit of vanishing diameter whereby the divergent parts of these quantities are explicitly calculated. It
is found that the divergent parts of the radial and azimuthal electric field components, as well as the electric potential, are
determined by the local charge density while the axial component is determined by the local dipole density.  For a straight
filament, these results reduce to those for conducting needles discussed in the literature. For curved filaments, the
configuration of charges and fields is no longer azimuthally symmetric, and there is an additional length scale in the problem
arising from the finite radius of curvature of the filament.  Remarkably, the basic uniformity result survives the added
complications, which include an azimuthal variation in the surface charge density of the filament.  As with the variations of
linear charge density along the filament, the azimuthal variations vanish with the characteristic parameter, only more rapidly.
These uniformity results yield an asymptotic formula for the capacitance of a curved filament that generalizes Maxwell's
original result.  The examples of a straight filament with uniform and linearly varying charge densities, as well as a circular filament
with a uniform charge distribution, are treated analytically and found to be in agreement with the results of the general
analysis.  Numerical calculations illustrating the slow convergence of linear charge distribution to uniformity for an
elliptical filament are presented, and an interactive computer program implementing and animating the numerical calculations is
provided.

\end{abstract}


\maketitle


\setlength{\arraycolsep}{0pt}
\section{introduction}
A paper of Griffiths and Li in this journal \cite{djgyli} has triggered a lively discussion on the electrostatic equilibrium distribution of charge on a
thin, straight, cylindrical conductor.  While suspecting that as the diameter-to-length ratio of the conductor approaches zero
the charge distribution on it tends to uniformity at all points sufficiently far from the ends, they encountered certain
puzzling aspects of the problem in their numerical studies.  In particular, they identified an apparent contradiction in the
strict limit of zero diameter, posing the question that ``\ldots if the charge density \textit{were} constant, on a truly one-dimensional needle,
how could the force on an off-center point be zero \ldots?''

Contributions by Good \cite{good} and Andrews \cite{and} followed the work of Griffiths and Li, offering further evidence of the
uniformity conjecture of the latter, as well as alternative methods of analyzing the problem. However, it fell to Jackson
\cite{jdj1} to provide a clear resolution of the equilibrium paradox quoted above by carefully considering the limit of
vanishing diameter-to-length ratio and the crucial role of the slow, inverse-logarithmic approach of charge density to
uniformity in that limit.

A subsequent paper by Jackson \cite{jdj2} reported on his discovery that the problem in question had already been considered by
none other than James Clerk Maxwell!   Maxwell, treating ``the electrical capacity of a long narrow cylinder'' more than 130
years ago \cite{jcm}, had derived the inverse-logarithmic approach to uniformity noted by Andrews and highlighted by Jackson,
and had eloquently summarized the behavior of electrostatic charges on a thin, straight conductor.  Maxwell's pioneering words
on the subject turned out to be a very fitting last word as well.

The purpose of this contribution is to investigate the distribution of charge on a thin, \textit{curved} conductor, which we
will often refer to as a \textit{filament}.  The intuition underlying this inquiry is the expectation that a sufficiently thin,
smoothly curved filament, when viewed up close, would appear long and straight, so that the equilibrium behavior of charges on
the filament should be nearly uniform just as in the case of straight conductors.  This is indeed what we find and describe in
this paper.

We approach the problem of charge distribution on a physical filament in two steps.  The first step is the treatment of the
fields of any sufficiently smooth, one-dimensional distribution of charge, which we will often refer to as an \textit{ideal filament}, with no regard to its
conductivity properties.  Here we consider the behavior of the electric field components and the electrostatic potential in the limit of vanishing
distance from an arbitrary point of the filament, and carry out an asymptotic analysis to calculate the divergent parts of these
quantities explicitly.  We find that the divergent parts of the radial and azimuthal electric field components, as well as the
electric potential, are determined by the linear charge density of the filament at the point in question, whereas the axial
component of the electric field is determined by the spacial derivative of linear charge density, or equivalently, the electric
dipole density, at the point.  The intuition underlying these results is discussed throughout, and the results are summarized in
Eqs.~(\ref{19}) and (\ref{20}).

In the second step of our analysis, we relate the fields of the ideal filament to those of a physical one by
casting the former as the symmetry axis of a curved, cylindrical conductor.  Here, we show that the equipotential surfaces of
the physical filament approach those of the ideal one as its diameter vanishes, and use this to deduce the inverse-logarithmic
approach of charge distribution to uniformity, the primary result of this paper, in Eq.~(\ref{24}).

The rest of this paper is organized as follows.  In \S IIA we set up the geometry of the ideal filament and formulate the
required regularity conditions.  In \S IIB we present the derivation of the electric field components near the ideal filament,
and in \S IIC we deal with the electrostatic potential.  We then formulate the requirements of relating the fields of the
physical filament to those of the ideal one in \S IIIA.  In \S IIIB we consider the consequences of enforcing those
requirements, whence we derive the result expressing the approach of charge density to uniformity.  An alternative derivation of
this result using energy minimization is developed in \S IIIC.  Examples of straight and circular filaments providing analytical
illustrations of Eqs.~(\ref{19}) and (\ref{20}) are given in \S IVA.  In \S IVB we present the results of a numerical study of
an elliptical filament, including an interactive computer program for calculating them.  Concluding remarks are presented in \S V.

\section{electric field components near an ideal charged filament}
\subsection{Geometry of a Regular Filament}
Consider a charged filament in the limit of zero diameter, i.e., idealized as an open or closed curve.  We will assume this
curve to be of class $\mathds{C}^2$, or twice continuously differentiable, and parameterized by $\mathbf{R}(s)$, ${s}_{1} \leq s
\leq {s}_{2} $, where $s$ is the arc length parameter \cite{smooth}. These regularity conditions guarantee the existence and
continuity of the principal normal vector and curvature, as well as the tangent vector, at every point.  Thus
$\boldsymbol{\hat{\tau}}(s)=d\mathbf{R}(s) /ds $ is the unit vector tangent to the curve at point $\mathbf{R}(s)$, with
$ds=(d\mathbf{R} \cdot d\mathbf{R})^{1/2}$.  Furthermore, the unit principal normal is given by
$\boldsymbol{\hat{\nu}}(s)={\kappa(s)}^{-1} d\boldsymbol{\hat{\tau}}(s)/ds$, where $\kappa(s)$ is the curvature at point $s$.
These two unit vectors, together with the unit \textit{binormal} vector
$\boldsymbol{\hat{\beta}}(s)=\boldsymbol{\hat{\tau}}(s)\times \boldsymbol{\hat{\nu}}(s)$, constitute a local orthogonal triad at
each point of the curve.  Note that the regularity conditions stated above guarantee that $\kappa(s)$ is a continuous function,
allowing us to define a radius of curvature $\varrho(s)=1/\kappa(s)$ which is continuous and has an infimum which we denote by
${\varrho}_{m}=1/{\kappa}_{m}$.  Note also that the same conditions imply that the curve is non-self-intersecting, i.e., that $s
\neq s' \Rightarrow \mathbf{R}(s) \neq \mathbf{R}(s')$, except for closed filaments for which $ \mathbf{R}({s}_{2}) =
\mathbf{R}({s}_{1})$.    The minimum radius of curvature and the length of the filament are the two relevant length scales for
points near the filament, and we will find it convenient to use the scale parameter $L=\min({\varrho}_{m},{s}_{2}-{s}_{1})$ in
the following analysis.

Consider a point ${P}_{\scriptscriptstyle 0}$ of the filament corresponding to ${s}_{\scriptscriptstyle 0}$, where ${s}_{1} \leq
{s}_{\scriptscriptstyle 0} \leq {s}_{2} $ for closed filaments and ${s}_{1} < {s}_{\scriptscriptstyle 0} < {s}_{2} $ in case of
open filaments.  As defined above, ${\boldsymbol{\hat{\tau}}}({s}_{\scriptscriptstyle 0})$,
$\boldsymbol{\hat{\nu}}({s}_{\scriptscriptstyle 0})$, and $\boldsymbol{\hat{\beta}}({s}_{\scriptscriptstyle 0})$ are the
tangent, normal, and binormal unit vectors at point ${P}_{\scriptscriptstyle 0}$.  We will also consider a nearby point $P$
located at $\mathbf{R}({s}_{\scriptscriptstyle 0})+\epsilon L {\mathbf{\hat{n}}}(\alpha,{s}_{\scriptscriptstyle 0})$ on the
normal plane of the filament at point ${P}_{\scriptscriptstyle 0}$, where ${\mathbf{\hat{n}}}({s}_{\scriptscriptstyle
0},\alpha)=\cos(\alpha){\boldsymbol{\hat{\nu}}}({s}_{\scriptscriptstyle
0})+\sin(\alpha){\boldsymbol{\hat{\beta}}}({s}_{\scriptscriptstyle 0}) $, $0 \leq \alpha \leq 2\pi$, is a unit vector in the
normal plane and $\epsilon L$ is the distance between ${P}_{\scriptscriptstyle 0}$ and $P$.  Here and below, $\epsilon$ is a
dimensionless, positive number which represents the distance ${P}_{\scriptscriptstyle 0}P$ in units of $L$.  Clearly, as
$\epsilon \rightarrow 0$, $P$ approaches ${P}_{\scriptscriptstyle 0}$.
\begin{figure}
\includegraphics[]{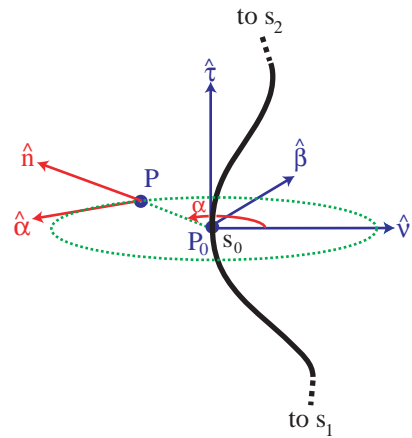}
\caption{The coordinate system used to describe the geometry of a curved, cylindrical filament.  The unit vectors ($\boldsymbol{\hat{\tau}},\boldsymbol{\hat{\nu}},\boldsymbol{\hat{\beta}}$) form the triad of tangent, normal, and binormal unit vectors at point ${P}_{\scriptscriptstyle 0}$ of the solid curve.  The pair ($\boldsymbol{\hat{\nu}},\boldsymbol{\hat{\beta}}$) defines the transverse plane at point ${P}_{\scriptscriptstyle 0}$, with the dotted circle delimiting the cross section of the curved cylinder.  The azimuthal angle $\alpha$ and the pair ($\boldsymbol{\hat{n}},\boldsymbol{\hat{\alpha}}$) define a polar coordinate system on the transverse plane.}
\label{fig1}
\end{figure}

Figure 1 illustrates the geometry of the filament and the coordinate system used to describe it. It also suggests that as
$\alpha$ ranges from $0$ to $2\pi$, the line ${P}_{\scriptscriptstyle 0}P$ sweeps the surface of a normal disk of radius
$\epsilon L$ centered on point ${P}_{\scriptscriptstyle 0}$. Relative to this disk,
${\mathbf{\hat{n}}}(\alpha,{s}_{\scriptscriptstyle 0})$ is the ``radial'' unit vector at point $P$ and
${\boldsymbol{\hat{\alpha}}}(\alpha,{s}_{\scriptscriptstyle 0})=-\sin(\alpha){\boldsymbol{\hat{\nu}}}({s}_{\scriptscriptstyle
0})+\cos(\alpha){\boldsymbol{\hat{\beta}}}({s}_{\scriptscriptstyle 0})$ the azimuthal unit vector.  These unit vectors, together
with ${\boldsymbol{\hat{\tau}}}({s}_{\scriptscriptstyle 0})$, constitute an orthogonal triad at point $P$ similar to the three
unit vectors of a cylindrical coordinate system.  Note that as $s$ ranges from ${s}_{1}$ to ${s}_{2}$ and $\alpha$ sweeps the
interval $[0,2\pi]$, the set of points $\mathbf{R}({s})+\epsilon L{\mathbf{\hat{n}}}(\alpha,{s})$ define the surface of a
curved, circular cylinder of fixed diameter $2\epsilon L$ \cite{tbp}.
\subsection{Electric Field Components}
The electric field at any point $\mathbf{r}$ not on the filament is given by
\begin{equation}
\mathbf{E}(\mathbf{r})=k\int_{{s}_{1}}^{{s}_{2}} ds \, \lambda({s})
[\mathbf{r}-\mathbf{R}({s})]/{|\mathbf{r}-\mathbf{R}({s})|}^{-3}, \label{1}
\end{equation}
where $\lambda({s})$ is the linear charge density on the filament, assumed to be continuously differentiable, and $k={(4\pi
{\epsilon}_{\scriptscriptstyle 0})}^{-1}$.  In particular, the electric field at point $P$ is given by
$\mathbf{E}[\mathbf{R}({s}_{\scriptscriptstyle 0})+\epsilon L {\mathbf{\hat{n}}}(\alpha,{s}_{\scriptscriptstyle 0})]$.

In the following we will be concerned with the axial, radial, and azimuthal components of the electric field at point
${P}_{\scriptscriptstyle 0}$, which we respectively define as the limit of ${E}_{\textrm{ax}}(P)$, ${E}_{\textrm{ra}}(P)$, and
${E}_{\textrm{az}}(P)$ as $P$ approaches ${P}_{\scriptscriptstyle 0}$.  We then find
\begin{eqnarray}
&&{E}_{\textrm{ax}}(\alpha,{s}_{\scriptscriptstyle 0})={\lim}_{\epsilon \rightarrow 0}
{\boldsymbol{\hat{\tau}}}({s}_{\scriptscriptstyle 0}) \cdot
\mathbf{E}[\mathbf{R}({s}_{\scriptscriptstyle 0})+\epsilon L {\mathbf{\hat{n}}}(\alpha,{s}_{\scriptscriptstyle 0})], \label{2} \\
&&{E}_{\textrm{\textrm{ra}}}(\alpha,{s}_{\scriptscriptstyle 0})={\lim}_{\epsilon \rightarrow 0} {\mathbf{\hat{n}}}(\alpha,{s}_{\scriptscriptstyle 0})\cdot \mathbf{E}[\mathbf{R}({s}_{\scriptscriptstyle 0})+\epsilon L {\mathbf{\hat{n}}}(\alpha,{s}_{\scriptscriptstyle 0})], \label{3} \\
&&{E}_{\textrm{\textrm{az}}}(\alpha,{s}_{\scriptscriptstyle 0})={\lim}_{\epsilon \rightarrow 0}
{\boldsymbol{\hat{\alpha}}}(\alpha,{s}_{\scriptscriptstyle 0})\cdot \mathbf{E}[\mathbf{R}({s}_{\scriptscriptstyle 0})+\epsilon L
{\mathbf{\hat{n}}}(\alpha,{s}_{\scriptscriptstyle 0})].\,\,\,\,\,\,\, \label{4}
\end{eqnarray}
Furthermore, we are only interested in the leading, divergent contributions to these limits, and will organize the following
calculations accordingly.  It should also be noted here that the electric field components at point ${P}_{\scriptscriptstyle 0}$
depend on the azimuthal angle $\alpha$ as well as on ${s}_{\scriptscriptstyle 0}$.  This is because the $\epsilon \rightarrow 0$
limit is a singular one, which in general renders the limit dependent on the azimuth of the approach to point
${P}_{\scriptscriptstyle 0}$.

Written in detail, Eqs.~(\ref{2})-(\ref{4}) appear as
\begin{eqnarray}
&&{E}_{\textrm{ax}}(\alpha,{s}_{\scriptscriptstyle 0})=k\, {\lim}_{\epsilon \rightarrow 0} \int_{{s}_{1}}^{{s}_{2}} ds \, \lambda(s){\boldsymbol{\hat{\tau}}}({s}_{\scriptscriptstyle 0})\nonumber \\
&&\cdot [\mathbf{R}({s}_{\scriptscriptstyle 0})-\mathbf{R}(s)]
{|\epsilon L {\mathbf{\hat{n}}}(\alpha,{s}_{\scriptscriptstyle 0}) +\mathbf{R}({s}_{\scriptscriptstyle 0})-\mathbf{R}(s)|}^{-3},\,\,\,\,\,\label{5}\\ &&{E}_{\textrm{ra}}(\alpha,{s}_{\scriptscriptstyle 0})=k\, {\lim}_{\epsilon \rightarrow 0}  \int_{{s}_{1}}^{{s}_{2}} ds \, \lambda(s)\{ \epsilon L + {\mathbf{\hat{n}}}(\alpha,{s}_{\scriptscriptstyle 0}) \nonumber \\
&&\cdot[\mathbf{R}({s}_{\scriptscriptstyle 0})-\mathbf{R}(s)] \} {|\epsilon L {\mathbf{\hat{n}}}(\alpha,{s}_{\scriptscriptstyle 0}) +\mathbf{R}({s}_{\scriptscriptstyle 0})-\mathbf{R}(s)|}^{-3},\,\,\,\,\,\label{6}\\ &&{E}_{\textrm{az}}(\alpha,{s}_{\scriptscriptstyle 0})=k\, {\lim}_{\epsilon \rightarrow 0}  \int_{{s}_{1}}^{{s}_{2}} ds \, \lambda(s) {\boldsymbol{\hat{\alpha}}}(\alpha,{s}_{\scriptscriptstyle 0})\nonumber \\ &&\cdot [\mathbf{R}({s}_{\scriptscriptstyle 0})-\mathbf{R}(s)]
{|\epsilon L {\mathbf{\hat{n}}}(\alpha,{s}_{\scriptscriptstyle 0})
+\mathbf{R}({s}_{\scriptscriptstyle 0})-\mathbf{R}(s)|}^{-3}.\,\,\,\,\,\label{7}
\end{eqnarray}
At this point it is convenient to use the regularity conditions imposed above on the charged filament, namely the $\mathds{C}^1$
and $\mathds{C}^2$ properties of $\lambda(s)$ and $\mathds{C}^2$ respectively \cite{regular}, to write
\begin{eqnarray}
\lambda(s)&=&\lambda({s}_{\scriptscriptstyle 0})+(s-{s}_{\scriptscriptstyle 0})\lambda'({s}_{\scriptscriptstyle 0})+\tilde{\lambda}(s,{s}_{\scriptscriptstyle 0}), \,\,\,\,\,\,\,\,\,\,\,\,\,\,\,\,\,\,\,\,\, \label{8} \\
\mathbf{R}(s)&=&\mathbf{R}({s}_{\scriptscriptstyle 0})+(s-{s}_{\scriptscriptstyle 0})
\boldsymbol{\hat{\tau}}({s}_{\scriptscriptstyle 0})+\frac{1}{2}{(s-{s}_{\scriptscriptstyle 0})}^{2}
\kappa({s}_{\scriptscriptstyle 0}) \boldsymbol{\hat{\nu}}({s}_{\scriptscriptstyle 0}) \nonumber \\
&+&\mathbf{\tilde{R}}(s,{s}_{\scriptscriptstyle 0}), \label{9}
\end{eqnarray}
where $\tilde{\lambda}(s,{s}_{\scriptscriptstyle 0})$ and $\mathbf{\tilde{R}}(s,{s}_{\scriptscriptstyle 0})$ and are
``remainder'' terms which vanish more rapidly than the terms preceding them as $s \rightarrow {s}_{\scriptscriptstyle 0}$.  In
the following, we will use the representations of Eqs.~(\ref{8}) and (\ref{9}) to organize each electric field component into
divergent (leading) and finite (remainder) parts as ${\epsilon \rightarrow 0}$.  Furthermore, we are primarily interested in the
divergent parts and will not explicitly calculate the finite remainder terms in the following.  We note in passing that the remainder terms would in general depend on the detailed structure of the filament and its charge distribution.

We will first consider the contributions to the radial component of the electric field near the filament,
${E}_{\textrm{ra}}^{(0)}(\alpha,{s}_{\scriptscriptstyle 0})$ and ${E}_{\textrm{ra}}^{(1)}(\alpha,{s}_{\scriptscriptstyle 0})$,
which arise from the first two terms in Eq.~(\ref{8}) respectively.  Let us first observe that $L$ (the smaller of filament
length and minimum radius of curvature) is the only physically relevant length scale as point ${P}_{\scriptscriptstyle 0}$ is
approached.  Clearly then, the filament should appear as a long, straight line of charge as viewed from any point
$\mathbf{R}({s}_{\scriptscriptstyle 0})+\epsilon L {\mathbf{\hat{n}}}(\alpha,{s}_{\scriptscriptstyle 0})$ if $\epsilon \ll 1$.
Consequently, we would intuitively expect ${E}_{\textrm{ra}}^{(0)}(P)$ to approach the electric field of an infinite, straight
line of constant charge density $\lambda({s}_{\scriptscriptstyle 0})$ through point ${P}_{\scriptscriptstyle 0}$ in the
direction of ${\boldsymbol{\hat{\tau}}}({s}_{\scriptscriptstyle 0})$ as $P \rightarrow {P}_{\scriptscriptstyle 0}$. In other
words, ${E}_{\textrm{ra}}^{(0)}(P)\rightarrow 2k \lambda({s}_{\scriptscriptstyle 0})/\epsilon L$ as $\epsilon \rightarrow 0$.
Similarly, we expect ${E}_{\textrm{ra}}^{(1)}(P)$ to approach the transverse field of a dipole of moment $a \lambda
'({s}_{\scriptscriptstyle 0}){\boldsymbol{\hat{\tau}}}({s}_{\scriptscriptstyle 0})$ centered at point ${P}_{\scriptscriptstyle
0}$, where $a$ is a proportionality constant (independent of $\epsilon$).  However, due to symmetry, the transverse field of a
dipole located at point ${P}_{\scriptscriptstyle 0}$ vanishes at point $P$.  Thus we expect the leading contribution to the
radial component of the electric field to equal ${E}_{\textrm{ra}}^{(0)}(\alpha,{s}_{\scriptscriptstyle 0})={\lim}_{\epsilon
\rightarrow 0} 2k \lambda({s}_{\scriptscriptstyle 0})/\epsilon L $.

To verify the above expectations for ${E}_{\textrm{ra}}(\alpha,{s}_{\scriptscriptstyle 0})$, we find it convenient to start by
making a change of variable from $s$ to $\xi =(s-{s}_{\scriptscriptstyle 0})/\epsilon L $.  The resulting expression can then be
written as
\begin{eqnarray}
&&{E}_{\textrm{ra}}^{(0)}(\alpha,{s}_{\scriptscriptstyle 0})= {\lim}_{\epsilon \rightarrow 0} [k \lambda({s}_{\scriptscriptstyle
0})/\epsilon L ]
\int_{{\Lambda}_{1} }^{{\Lambda}_{2}}d\xi \nonumber \\
&&\times \big \{ 1 +\xi {\mathbf{\hat{n}}}(\alpha,{s}_{\scriptscriptstyle 0})
\cdot [\mathbf{R}({s}_{\scriptscriptstyle 0})-  \mathbf{R}({s}_{\scriptscriptstyle 0}+\epsilon L \xi)] /\epsilon L \xi  \big \}\nonumber \\
&&\times {\big|{\mathbf{\hat{n}}}(\alpha,{s}_{\scriptscriptstyle 0})+\xi [\mathbf{R}({s}_{\scriptscriptstyle 0})-
\mathbf{R}({s}_{\scriptscriptstyle 0}+\epsilon L \xi)]/\epsilon L \xi \big |}^{-3}, \label{10}
\end{eqnarray}
where ${\Lambda}_{1}=[{s}_{1}-{s}_{\scriptscriptstyle 0}]/\epsilon L$ and  ${\Lambda}_{2}=[{s}_{2}-{s}_{\scriptscriptstyle
0}]/\epsilon L$. Recalling from Eq.~(\ref{9}) that $[\mathbf{R}({s}_{\scriptscriptstyle 0}+\epsilon L
\xi)-\mathbf{R}({s}_{\scriptscriptstyle 0})]/\epsilon L \xi \rightarrow \mathbf{R}'({s}_{\scriptscriptstyle 0})+\frac{1}{2}\epsilon L \xi\mathbf{R}''({s}_{\scriptscriptstyle 0})=
{\boldsymbol{\hat{\tau}}}({s}_{\scriptscriptstyle 0})+\frac{1}{2}\epsilon L \xi \kappa({s}_{\scriptscriptstyle 0}) \boldsymbol{\hat{\nu}}({s}_{\scriptscriptstyle 0})$ as $\epsilon \rightarrow 0$, we can rewrite the above result as
\begin{eqnarray}
{E}_{\textrm{ra}}^{(0)}(\alpha,{s}_{\scriptscriptstyle 0})&=&{\lim}_{\epsilon \rightarrow 0} [k \lambda({s}_{\scriptscriptstyle
0})/\epsilon L] \int_{{\Lambda}_{1}}^{{\Lambda}_{2}}  d\xi [1-\frac{1}{2}\epsilon L \kappa({s}_{\scriptscriptstyle 0})\nonumber \\
&\times& \cos(\alpha){\xi}^2]     {| {\mathbf{\hat{n}}}(\alpha,{s}_{\scriptscriptstyle 0})-\xi
{\boldsymbol{\hat{\tau}}}({s}_{\scriptscriptstyle 0})|}^{-3}. \label{11}
\end{eqnarray}
The first term in the above integral is readily found to equal $2$ by noting that $| {\mathbf{\hat{n}}}(\alpha,{s}_{\scriptscriptstyle 0}) -\xi{\boldsymbol{\hat{\tau}}}({s}_{\scriptscriptstyle 0})|={(1+ {\xi}^{2})}^{1/2}$ and that ${\Lambda}_{(1,2)}\rightarrow (-\infty,+\infty)$ in the limit. We therefore find that the first term in Eq.~(\ref{11}) is given by ${\lim}_{\epsilon \rightarrow 0} 2k \lambda({s}_{\scriptscriptstyle
0})/\epsilon L $, confirming our intuitive expectation above.

The second term in Eq.~(\ref{11}) is a correction to the leading contribution we anticipated, and involves the integral $\int_{{\Lambda}_{1}}^{{\Lambda}_{2}}  d\xi\, {\xi}^2 {(1+ {\xi}^{2})}^{-3/2}$.  The asymptotic value of this integral is found to equal $\ln({\epsilon}^{-2})$ by noting that ${\Lambda}_{1}$ and ${\Lambda}_{2}$ are both of the order of ${\epsilon}^{-1}$ as $\epsilon \rightarrow 0$.  Using this result, we find that the second term in Eq.~(\ref{11}) is given by $k  \lambda({s}_{\scriptscriptstyle 0})\kappa({s}_{\scriptscriptstyle 0})\cos(\alpha)
{\lim}_{\epsilon \rightarrow 0}\ln(\epsilon)$.

Combining the two contributions from Eq.~(\ref{11}), we find
\begin{equation}
{E}_{\textrm{ra}}^{(0)}(\alpha,{s}_{\scriptscriptstyle 0})={\lim}_{\epsilon \rightarrow 0} k \lambda({s}_{\scriptscriptstyle
0})[2{(\epsilon L)}^{-1}+\kappa({s}_{\scriptscriptstyle 0})\cos(\alpha) \ln(\epsilon)]. \label{12}
\end{equation}
It is worth pointing out here that just as we anticipated the leading term in Eq.~(\ref{12}) as corresponding to a long,
straight line of charge, we could have arrived at the second contribution by noting that, in the second approximation, the
filament can be considered a circular arc of radius $1/\kappa({s}_{\scriptscriptstyle 0})$, i.e., the radius of curvature of the
filament at the point in question.  The reader can verify this by reference to the treatment of the circular filament in \S IVA.

Turning to ${E}_{\textrm{ra}}^{(1)}(\alpha,{s}_{\scriptscriptstyle 0})$ next, and using a procedure similar to that followed
above for ${E}_{\textrm{ra}}^{(0)}(\alpha,{s}_{\scriptscriptstyle 0})$, we find
\begin{eqnarray}
&&{E}_{\textrm{ra}}^{(1)}(\alpha,{s}_{\scriptscriptstyle 0})=k\,{\lim}_{\epsilon \rightarrow 0} \int_{{\Lambda}_{1}
}^{{\Lambda}_{2}} \xi d\xi \, \frac{\lambda({s}_{\scriptscriptstyle 0}+\epsilon L \xi) - \lambda({s}_{\scriptscriptstyle
0})}{\epsilon L \xi}
\nonumber \\
&&\times \big \{ 1 +\xi {\mathbf{\hat{n}}}(\alpha,{s}_{\scriptscriptstyle 0})
\cdot [\mathbf{R}({s}_{\scriptscriptstyle 0})-  \mathbf{R}({s}_{\scriptscriptstyle 0}+\epsilon L \xi)] /\epsilon L \xi  \big \}\nonumber \\
&&\times {\big|{\mathbf{\hat{n}}}(\alpha,{s}_{\scriptscriptstyle 0})+\xi [\mathbf{R}({s}_{\scriptscriptstyle 0})-
\mathbf{R}({s}_{\scriptscriptstyle 0}+\epsilon L \xi)]/\epsilon L \xi \big |}^{-3}.\,\,\,\,\,\, \label{13}
\end{eqnarray}
Using Eqs.~(\ref{8}) and (\ref{9}), we implement the $\epsilon \rightarrow 0$ limit in the above expression to find
\begin{equation}
{E}_{\textrm{ra}}^{(1)}(\alpha,{s}_{\scriptscriptstyle 0})=k \lambda'({s}_{\scriptscriptstyle 0}) \int_{-\infty}^{+ \infty}\xi
d\xi {| {\mathbf{\hat{n}}}(\alpha,{s}_{\scriptscriptstyle 0})-\xi {\boldsymbol{\hat{\tau}}}({s}_{\scriptscriptstyle 0})|}^{-3} ,
\label{14}
\end{equation}
which vanishes by symmetry as anticipated.  We have thus verified that ${E}_{\textrm{ra}}(\alpha,{s}_{\scriptscriptstyle 0})
\rightarrow {\tilde{E}}_{\textrm{ra}}(\alpha,{s}_{\scriptscriptstyle 0})+ 2k \lambda({s}_{\scriptscriptstyle 0})/\epsilon L$ as
$\epsilon \rightarrow 0$, where ${\tilde{E}}_{\textrm{ra}}(\alpha,{s}_{\scriptscriptstyle 0})$ is the finite part of the radial
electric field at point ${P}_{\scriptscriptstyle 0}$.

Our next task is to deal with the axial component of the electric field, ${E}_{\textrm{ax}}(\alpha,{s}_{\scriptscriptstyle 0})$.
For reasons of symmetry, we would expect ${E}_{\textrm{ax}}^{(0)}(\alpha,{s}_{\scriptscriptstyle 0})$, the contribution arising
from $\lambda({s}_{\scriptscriptstyle 0})$ in (\ref{8}), to vanish since it corresponds to the axial electric field of a
straight filament of uniform charge density as the filament is approached. Indeed following a procedure parallel to that used
above, we find that ${E}_{\textrm{ax}}^{(0)}(\alpha,{s}_{\scriptscriptstyle 0})$ is given by
\begin{eqnarray}
{E}_{\textrm{ax}}^{(0)}(\alpha,{s}_{\scriptscriptstyle 0})&=&-k \lambda({s}_{\scriptscriptstyle 0})\, {\lim}_{\epsilon
\rightarrow 0} {(\epsilon L)}^{-1} \int_{-\infty}^{+\infty}  \xi d\xi \nonumber \\ &\times&
{|{\mathbf{\hat{n}}}(\alpha,{s}_{\scriptscriptstyle 0})-\xi {\boldsymbol{\hat{\tau}}}({s}_{\scriptscriptstyle 0})|}^{-3},
\label{15}
\end{eqnarray}
which vanishes by symmetry as surmised.

The second contribution to the axial component of the electric field, the one arising from $(s-{s}_{\scriptscriptstyle
0})\lambda'({s}_{\scriptscriptstyle 0})$ in (\ref{8}), corresponds to the longitudinal field of an electric dipole of moment $a
\lambda '({s}_{\scriptscriptstyle 0}){\boldsymbol{\hat{\tau}}}_{\scriptscriptstyle 0}$ centered at point
${P}_{\scriptscriptstyle 0}$, as stated earlier. Indeed following the procedure established above, we find
\begin{eqnarray}
{E}_{\textrm{ax}}^{(1)}(\alpha,{s}_{\scriptscriptstyle 0})=-k\lambda'({s}_{\scriptscriptstyle 0})\, {\lim}_{\epsilon \rightarrow
0}  \int_{{\Lambda}_{1} }^{{\Lambda}_{2}} {\xi}^{2} d\xi \nonumber \\
\times {|{\mathbf{\hat{n}}}(\alpha,{s}_{\scriptscriptstyle 0})-\xi {\boldsymbol{\hat{\tau}}}({s}_{\scriptscriptstyle 0})|}^{-3}
, \label{16}
\end{eqnarray}
which can be evaluated by using the fact that $|\mathbf{\hat{n}}(\alpha,{s}_{\scriptscriptstyle 0})-\xi
{\boldsymbol{\hat{\tau}}}({s}_{\scriptscriptstyle 0})|={(1+{\xi}^{2})}^{1/2}$ as noted above.  The result is
\begin{equation}
{E}_{\textrm{ax}}^{(1)}(\alpha,{s}_{\scriptscriptstyle 0})={\lim}_{\epsilon \rightarrow 0} 2k \lambda'({s}_{\scriptscriptstyle
0}) \ln(\epsilon), \label{17}
\end{equation}
a logarithmically divergent expression.  We have thus found that the axial component of the electric field
${E}_{\textrm{ax}}(\alpha,{s}_{\scriptscriptstyle 0}) \rightarrow {\tilde{E}}_{\textrm{ax}}(\alpha,{s}_{\scriptscriptstyle 0})+
2k \lambda'({s}_{\scriptscriptstyle 0}) \ln(\epsilon)$ as $\epsilon \rightarrow 0$, where
${\tilde{E}}_{\textrm{ax}}(\alpha,{s}_{\scriptscriptstyle 0})$ is the finite part of the axial electric field at point
${P}_{\scriptscriptstyle 0}$.

The calculation of the azimuthal component ${E}_{\textrm{az}}(\alpha,{s}_{\scriptscriptstyle 0})$ proceeds along similar lines,
with ${E}_{\textrm{az}}^{(0)}(\alpha,{s}_{\scriptscriptstyle 0})$ reducing to
\begin{equation}
{E}_{\textrm{az}}^{(0)}(\alpha,{s}_{\scriptscriptstyle 0})= {\lim}_{\epsilon \rightarrow 0}\, k \lambda({s}_{\scriptscriptstyle
0})\kappa({s}_{\scriptscriptstyle 0})\sin(\alpha) \ln({\epsilon}^{-1}), \label{18}
\end{equation}
and ${E}_{\textrm{az}}^{(1)}(\alpha,{s}_{\scriptscriptstyle 0})$ vanishing by symmetry.  Note that unlike the other components,
the azimuthal component of the electric field arises from the local curvature of the filament, and as such depends on the
azimuthal angle $\alpha$.   We therefore have the result that the azimuthal  component of the electric field
${E}_{\textrm{az}}(\alpha,{s}_{\scriptscriptstyle 0}) \rightarrow {\tilde{E}}_{\textrm{az}}(\alpha,{s}_{\scriptscriptstyle 0})+
k \lambda({s}_{\scriptscriptstyle 0})\kappa({s}_{\scriptscriptstyle 0})\sin(\alpha) \ln({\epsilon}^{-1})$ as $\epsilon
\rightarrow 0$, where ${\tilde{E}}_{\textrm{az}}(\alpha,{s}_{\scriptscriptstyle 0})$ is the finite part of the azimuthal
electric field at point ${P}_{\scriptscriptstyle 0}$.

At this point it is convenient to summarize the above results for the electric field components near a charged filament:

\textit{The electric field at an interior point ${P}_{\scriptscriptstyle 0}$ of a charged, regular, ideal filament is given by}
\begin{eqnarray}
&&\mathbf{E}(\alpha,{s}_{\scriptscriptstyle 0})=\mathbf{\tilde{E}}(\alpha,{s}_{\scriptscriptstyle 0})+k\,{\lim}_{\epsilon
\rightarrow 0} \bigl\{ 2\lambda'({s}_{\scriptscriptstyle 0}) \ln(\epsilon ){\boldsymbol{\hat{\tau}}}({s}_{\scriptscriptstyle 0})
\nonumber \\&& +  \lambda({s}_{\scriptscriptstyle 0})[2{(\epsilon L)}^{-1}+\kappa({s}_{\scriptscriptstyle 0})\cos(\alpha)
\ln(\epsilon)] {\mathbf{\hat{n}}}(\alpha,{s}_{\scriptscriptstyle 0})\nonumber \\&&  + \lambda({s}_{\scriptscriptstyle 0})
\sin(\alpha)\kappa({s}_{\scriptscriptstyle 0}) \ln({\epsilon }^{-1}) \boldsymbol{\hat{\alpha}} (\alpha,{s}_{\scriptscriptstyle
0}) \bigr\} ,\, \label{19}
\end{eqnarray}
\textit{where} $\mathbf{\tilde{E}}(\alpha,{s}_{\scriptscriptstyle 0})$ \textit{is the finite part of the electric field at
point} ${P}_{\scriptscriptstyle 0}$ \textit{and the notation is as established above}.

It is worth repeating here that the divergent parts of the electric field in Eq.~(\ref{19}) arise from the local charge and
dipole densities at point ${P}_{\scriptscriptstyle 0}$. This is of course what is expected physically since, given the
regularity requirements on the geometry and distribution of charge on the filament \cite{regular}, there is no other source of
divergence than the proximity of the charges to the field point.  More importantly, while divergent parts of the radial and
azimuthal electric field components are generated by the local charge density, the divergent part of the axial component is
generated by the local dipole density. It is the latter feature of the fields and sources that will ensure the vanishing of the
dipole density (hence approach to uniformity) for a conducting filament in the limit of vanishing diameter, as the argument
leading to Eq.~(\ref{24}) will show. It should also be emphasized here that the results of this section apply to an ideal
filament of charge regardless of its conductivity properties.
\subsection{Electrostatic Potential}
We conclude this section by deriving the electrostatic potential near the filament.  The divergent part of this potential is
expected to correspond to a long, straight filament of constant charge density.  Indeed applying the methods established above,
we find the potential at point ${P}_{\scriptscriptstyle 0}$ of the filament as
\begin{equation}
\Phi(\alpha,{s}_{\scriptscriptstyle 0})=\tilde{\Phi}(\alpha,{s}_{\scriptscriptstyle 0})+2k \,{\lim}_{\epsilon \rightarrow 0}
\lambda({s}_{\scriptscriptstyle 0}) \ln({\epsilon }^{-1}), \label{20}
\end{equation}
where $\tilde{\Phi}(\alpha,{s}_{\scriptscriptstyle 0})$ is the finite part of the potential at point ${P}_{\scriptscriptstyle
0}$.  As expected, the divergent part of this result is just the electrostatic potential of a straight filament of charge
density $\lambda({s}_{\scriptscriptstyle 0})$, and is independent of the curvature of the filament at point
${P}_{\scriptscriptstyle 0}$.  As such, it does not contain any information on the azimuthal component of the electric field, or
the subleading contribution to the radial component, both of which are present in Eq.~(\ref{19}), while the axial and the
leading radial components are readily derivable from it, as will be shown below.  The potential terms corresponding to the
missing components, being of the order of $\epsilon \ln(\epsilon)$, are finite and vanish in the ${\epsilon \rightarrow 0}$
limit, which is the reason they don't appear in Eq.~(\ref{20}).

To extract the radial and axial electric field components from the potential, we note that the variables
${s}_{\scriptscriptstyle 0}$, $\epsilon $, and $\alpha$ on which the terms on the right-hand side of Eq.~(\ref{20}) depend are
in effect the axial, (rescaled) radial, and azimuthal coordinates of a local cylindrical coordinate system.  Consequently, we
can write
\begin{eqnarray}
&&\nabla \Phi(\alpha,{s}_{\scriptscriptstyle 0})=\nabla \tilde{\Phi}(\alpha,{s}_{\scriptscriptstyle 0})+2k \,{\lim}_{\epsilon \rightarrow 0} \nonumber \\
&\times&[{\boldsymbol{\hat{\tau}}}({s}_{\scriptscriptstyle 0})\partial/\partial {s}_{\scriptscriptstyle 0} +
{\mathbf{\hat{n}}}(\alpha,{{s}_{\scriptscriptstyle 0}}){L}^{-1}\partial/\partial \epsilon] \lambda({{s}_{\scriptscriptstyle 0}})
\ln({\epsilon }^{-1}). \label{21}
\end{eqnarray}
Note that while $\tilde{\Phi}(\alpha,{s}_{\scriptscriptstyle 0})$ is finite as ${\epsilon \rightarrow 0}$, its gradient is not
and in fact includes a logarithmically divergent azimuthal part as mentioned above.

Implementing the derivatives in Eq.~(\ref{21}), we find
\begin{eqnarray}
&&\nabla \Phi(\alpha,{s}_{\scriptscriptstyle 0})=\nabla \tilde{\Phi}(\alpha,{s}_{\scriptscriptstyle 0})-2k \, {\lim}_{\epsilon \rightarrow 0} \nonumber \\
&\times&[{\boldsymbol{\hat{\tau}}}({s}_{\scriptscriptstyle 0})\lambda'({s}_{\scriptscriptstyle 0}) \ln({\epsilon }) +
{\mathbf{\hat{n}}}(\alpha,{s}_{\scriptscriptstyle 0})\lambda({s}_{\scriptscriptstyle 0}) {({\epsilon }L)}^{-1}]. \label{22}
\end{eqnarray}
Note that $-\nabla {\Phi}(\alpha,{s}_{\scriptscriptstyle 0})\neq \mathbf{{E}}(\alpha,{s}_{\scriptscriptstyle 0})$, the
difference being the two $\alpha$-dependent terms in Eq.~(\ref{19}) which are absent in Eq.~(\ref{20}), as anticipated earlier.
Taking this into account, we find that Eqs.~(\ref{22}) and (\ref{19}) are in agreement with respect to the leading radial and
axial components.  Since the approach to a uniform charge density for a thin conductor found in Eq.~(\ref{24}) below is forged
from these leading components, we are assured that the electrostatic potential contains sufficient information to imply the same
result.  This is an important point, since the derivation of the uniformity result in \S IIIC  is based on energy considerations
using the electrostatic potential.

Anticipating the uniformity of charge distribution on a conducting filament in the limit of vanishing diameter, as expressed in
Eq.~(\ref{24}) below, we can use the asymptotic behavior given in Eq.~(\ref{20}) to derive a universal asymptotic formula for
the capacitance per unit length of a conducting filament. For such a filament, charge per unit length tends to a constant, say
$\lambda$, and the electrostatic potential to $2k \lambda \ln({\epsilon }^{-1})$, as seen from Eq.~(\ref{20}). Therefore the
capacitance per unit length, $c(\epsilon)$, tends to
\begin{equation}
c(\epsilon)=[2k \ln(1/{\epsilon })]^{-1}, \label{23}
\end{equation}
where, it may be recalled from \S II, $2\epsilon L$ is the diameter of the filament.  For a straight filament, Eq.~(\ref{23})
reduces to the result first derived by Maxwell \cite{jcm}; cf. \cite{jdj1}.

\section{distribution of charge on thin conducting filaments}
\subsection{Physical versus Ideal Filament}
We are now in a position to consider the physical problem of a charged, conducting filament in the limit of vanishing diameter.
We will define the physical filament by reference to the equipotential surfaces of the idealized charge distribution treated in
the previous section \cite{lcdev}.  Specifically, given a value ${\Phi}_{\scriptscriptstyle 0}$ for the potential of the
conductor (with respect to infinity), we consider the equipotential surface
$\textsf{S}[\mathbf{R}(\cdot),\lambda(\cdot),{\Phi}_{\scriptscriptstyle 0}]$ consisting of all points $\mathbf{r}$ such that
$\Phi(\mathbf{r})={\Phi}_{\scriptscriptstyle 0}$ to be the surface of the conducting, physical filament.  Here
$\Phi(\mathbf{r})$ is the (exact) electrostatic potential produced by the idealized filament.  The uniqueness property of the
Dirichlet problem would then assure us that the electrostatic field produced by $\lambda(s)$ in the exterior of $\textsf{S}$ is
precisely the same as would be produced by the physical conductor if held at potential ${\Phi}_{\scriptscriptstyle 0}$.
Furthermore, the surface charge per unit length of the physical conductor would be given by $\lambda(s)$ \cite{lcd}.

Having defined the physical conductor by reference to an assumed charge distribution, we have shifted the burden of our analysis
to reconciling the resulting profile with the actual shape of the physical conductor.  In particular, we might wonder if, for a
given $\mathbf{R}(s)$, it is possible to choose $\lambda(s)$ so that $\textsf{S}[\mathbf{R}(\cdot), \lambda(\cdot),
{\Phi}_{\scriptscriptstyle 0}]$ has a desired form, e.g., that of a curved, circular cylinder of fixed diameter defined in \S
IIA.  Here we have the benefit of the insight gained in the study of straight filaments, especially in Refs.
\cite{djgyli,and,jdj1,rowley}, where we find an affirmative answer.  In particular, we know that a conductor with an ellipsoidal
surface corresponds to a uniform charge distribution along the symmetry axis \cite{wrs}.  We also know, starting with Maxwell
\cite{jcm}, that a thin, cylindrical conductor acquires a near-uniform charge distribution except for endpoints where higher
charge densities occur.  Note that the field and charge configurations of a straight filament are azimuthally symmetric.

In the case of curved filaments of finite diameter, however, there's no such symmetry and one cannot in general expect a one-dimensional charge distribution to properly reproduce the azimuthal distribution of the charges on the filament's surface.
This shortcoming is of course a consequence of the mismatch between the dimensionality of an ideal filament and that of the
surface of a conductor.  By the same token, however, any difference between the two is expected to be small for a thin filament
and tend to zero as its diameter vanishes.  This expectation is in fact born out as the analysis below shows, and we will be
able to determine the linear charge density of a sufficiently thin conductor regardless of its azimuthal variations.

\subsection{Boundary Conditions}
The standard electrostatic boundary condition on a conductor is the uniformity of the potential throughout its interior and
surface, a condition which follows from the vanishing of the electric field in the interior of a conductor in electrostatic
equilibrium.  Our problem then is to constrain $\lambda(s)$ such that $\textsf{S}[\mathbf{R}(\cdot) , \lambda(\cdot),
{\Phi}_{\scriptscriptstyle 0}]$ turns out to be the curved, circular cylinder of fixed diameter defined above.  Equivalently, we
will require the vanishing of the tangential components of the electric field on the surface of the conductor, starting with
the axial component.

The axial component was calculated in \S IIB where we found that ${E}_{\textrm{ax}}(\alpha,{s}_{\scriptscriptstyle 0})
\rightarrow {\tilde{E}}_{\textrm{ax}}(\alpha,{s}_{\scriptscriptstyle 0})+ 2k \lambda'({s}_{\scriptscriptstyle 0}) \ln(\epsilon)$
as ${\lim}_{\epsilon \rightarrow 0}$.  The requirement that ${E}_{\textrm{ax}}(\alpha,{s}_{\scriptscriptstyle 0})$ vanish in the
limit immediately constrains the charge distribution according to
\begin{equation}
\lambda'({s}_{\scriptscriptstyle 0})=-{\lim}_{\epsilon \rightarrow 0}{\tilde{E}}_{\textrm{ax}}(\alpha,{s}_{\scriptscriptstyle
0})/2k \ln(\epsilon). \label{24}
\end{equation}
Recalling that $\mathbf{R}({s}_{\scriptscriptstyle 0})$ is an arbitrary point of the filament (with the exception of endpoints
in case of open filaments), we conclude that, as its diameter tends to zero, the linear charge density on the filament tends to
uniformity as the inverse of the logarithm of $\epsilon$.  This result generalizes and closely parallels those found for
straight filaments, starting with Maxwell's \cite{jcm}.

Next, we consider the azimuthal component of the electric field on the surface of the filament.  At first blush, Eq.~(\ref{19})
seems to imply that the vanishing of the azimuthal component requires the eventual disappearance of all charges on the filament!
In reality, however, what is conveyed by Eq.~(\ref{19}) is simply that the equipotential surface $\textsf{S}[\mathbf{R}(\cdot),
\lambda(\cdot), {\Phi}_{\scriptscriptstyle 0}]$ generated by an ideal filament cannot have a \textit{circular} cross section
except at points where the curvature of the filament vanishes.  This is of course exactly what is expected, since the deviation
from circularity is needed to balance the electric field component generated by the local curvature of the filament.  This sort
of behavior is already familiar from the study of straight conductors where variations in the diameter along the conductor are
required to balance the effect of unequal distances from the ends, and has been discussed in detail in Ref.~\cite{jdj2}.
Needless to say, this circumstance does not imply that circular cross-section filaments are unphysical, only that our
one-dimensional filament within is not capable of masquerading them.  Fortunately, it turns out that the deviation from
circularity is quite small and disappears rapidly as ${\epsilon \rightarrow 0}$.  This is the justification for our use of an
ideal charge filament to describe the distribution of charge on a thin, physical filament.

To establish the above assertion, we will examine the shape of the non-circular cross section generated by the ideal filament.
The unit normal ${\mathbf{\hat{N}}}(\alpha,{s}_{\scriptscriptstyle 0})$ to the equipotential surface
$\textsf{S}[\mathbf{R}(\cdot), \lambda(\cdot), {\Phi}_{\scriptscriptstyle 0}]$ at point $\mathbf{R}({s}_{\scriptscriptstyle
0})+\epsilon L {\mathbf{\hat{n}}}(\alpha,{s}_{\scriptscriptstyle 0})$ is defined by the direction of the electric field in Eq.~(\ref{19}), i.e.,
\begin{eqnarray}
&&{\mathbf{\hat{N}}}(\alpha,{s}_{\scriptscriptstyle 0})\propto {\lim}_{\epsilon \rightarrow
0}\,\big[\lambda'({s}_{\scriptscriptstyle 0}) \ln(\epsilon ){\boldsymbol{\hat{\tau}}}({s}_{\scriptscriptstyle 0})+
\lambda({s}_{\scriptscriptstyle 0})  {(\epsilon L)}^{-1} \nonumber \\ && \times{\mathbf{\hat{n}}}(\alpha,{s}_{\scriptscriptstyle
0}) +\frac{1}{2} \sin(\alpha)\lambda({s}_{\scriptscriptstyle 0})  \kappa({s}_{\scriptscriptstyle 0}) \ln({\epsilon }^{-1})
\boldsymbol{\hat{\alpha}} (\alpha,{s}_{\scriptscriptstyle 0})\big],\,\,\,\,\,\, \label{25}
\end{eqnarray}
where we have only included the leading contribution in each component.  Simplifying further, we find
\begin{eqnarray}
{\mathbf{\hat{N}}}(\alpha,{s}_{\scriptscriptstyle 0})&=&{\lim}_{\epsilon \rightarrow 0}\bigl\{
{\mathbf{\hat{n}}}(\alpha,{s}_{\scriptscriptstyle 0})+[L\lambda'({s}_{\scriptscriptstyle 0})/\lambda({s}_{\scriptscriptstyle
0})]\epsilon \ln(\epsilon ){\boldsymbol{\hat{\tau}}}({s}_{\scriptscriptstyle 0})\nonumber \\ &+&
\frac{1}{2}\kappa({s}_{\scriptscriptstyle 0})L \epsilon \ln({\epsilon }^{-1}) \sin(\alpha)\boldsymbol{\hat{\alpha}}
(\alpha,{s}_{\scriptscriptstyle 0}) \bigr\}. \label{26}
\end{eqnarray}

Let $\epsilon=\eta(\alpha,{s}_{\scriptscriptstyle 0},{\Phi}_{\scriptscriptstyle 0})$ describe the equipotential surface
$\textsf{S}[\mathbf{R}(\cdot), \lambda(\cdot), {\Phi}_{\scriptscriptstyle 0}]$ in terms of our curved cylindrical coordinates
$(\epsilon, \alpha, {s}_{\scriptscriptstyle 0})$.  Since the gradient vector $\nabla
[\epsilon-\eta(\alpha,{s}_{\scriptscriptstyle 0},{\Phi}_{\scriptscriptstyle 0})]$ is normal to the surface in question, its
direction must be the same as the unit normal  ${\mathbf{\hat{N}}}(\alpha,{s}_{\scriptscriptstyle 0})$.  Therefore, using
Eq.~(\ref{26}), we can write
\begin{eqnarray}
&&\partial\ln[\eta(\alpha,{s}_{\scriptscriptstyle 0},{\Phi}_{\scriptscriptstyle 0})]/ \partial \alpha =
\frac{1}{2}\kappa({s}_{\scriptscriptstyle 0})L \nonumber \\ &&\times {\eta(\alpha,{s}_{\scriptscriptstyle
0},{\Phi}_{\scriptscriptstyle 0})} \ln[{\eta(\alpha,{s}_{\scriptscriptstyle 0},{\Phi}_{\scriptscriptstyle 0})}]
\sin(\alpha),\,\,\,\,\,\,\,\,\,\,\,  \label{27}
\end{eqnarray}
which is valid to leading order as ${\lim}_{{\Phi}_{\scriptscriptstyle 0} \rightarrow \infty}$, or equivalently, as
$\eta(\alpha,{s}_{\scriptscriptstyle 0},{\Phi}_{\scriptscriptstyle 0}) \rightarrow 0$.  It is clear from Eq.~(\ref{27}), and can
be readily verified numerically, that the maximum fractional variation in the diameter of the equipotential surface with respect
to $\alpha$ is of the order of ${\eta(\alpha,{s}_{\scriptscriptstyle 0},{\Phi}_{\scriptscriptstyle 0})}
\ln[{\eta(\alpha,{s}_{\scriptscriptstyle 0},{\Phi}_{\scriptscriptstyle 0})}]$, hence quite small for a sufficiently small
diameter \cite{recall}.

It is important to emphasize here that the equipotential surface $\epsilon=\eta(\alpha,{s}_{\scriptscriptstyle
0},{\Phi}_{\scriptscriptstyle 0})$ is  normal to the electric field by construction, so that the tangential component of the
electric field vanishes identically at all points of this surface.  Consequently, the relevant boundary condition in matching
this equipotential surface to the surface of a curved, cylindrical conductor is the circularity of the transverse boundary of
the former, i.e., the vanishing of the partial derivative $\partial  \ln[\sigma(\alpha,{s}_{\scriptscriptstyle
0},{\Phi}_{\scriptscriptstyle 0})] /\partial \alpha$.  This is of course what is guaranteed by Eq.~(\ref{27}) in the limit of
vanishing diameter.

One can also examine the azimuthal variations of the surface charge density on the equipotential surface.  The latter is given
by the standard formula ${\epsilon}_{\scriptscriptstyle 0}{\mathbf{\hat{N}}}(\alpha,{s}_{\scriptscriptstyle 0}) \cdot
\mathbf{E}(\alpha,{s}_{\scriptscriptstyle 0})$.  Using Eqs.~(\ref{19}) and (\ref{27}), we find for the surface charge density,
\begin{eqnarray}
&& \sigma(\alpha,{s}_{\scriptscriptstyle 0},{\Phi}_{\scriptscriptstyle 0}) \cong  \frac{\lambda({s}_{\scriptscriptstyle 0})}{2 \pi L \eta(\alpha,{s}_{\scriptscriptstyle 0},{\Phi}_{\scriptscriptstyle 0}) }
\Bigl\{ 1 + \frac{1}{2} \nonumber \\
&& \times {\bigl[ \frac{1}{2}\kappa({s}_{\scriptscriptstyle 0})L \eta(\alpha,{s}_{\scriptscriptstyle
0},{\Phi}_{\scriptscriptstyle 0}) \ln[{\eta(\alpha,{s}_{\scriptscriptstyle 0},{\Phi}_{\scriptscriptstyle 0})} ] \sin(\alpha)
\bigr] }^{2} \Bigr\}, \,\,\,\,\, \label{28}
\end{eqnarray}
where we have kept the first two terms in $\eta(\alpha,{s}_{\scriptscriptstyle 0},{\Phi}_{\scriptscriptstyle 0})$ and suppressed the dependence of the linear charge density $\lambda({s}_{\scriptscriptstyle 0})$ on ${\Phi}_{\scriptscriptstyle 0}$.

The quantity of interest here is $\partial \ln[\sigma(\alpha,{s}_{\scriptscriptstyle 0},{\Phi}_{\scriptscriptstyle 0})]
/\partial \alpha$, the fractional variation of surface charge density with respect to the azimuthal angle.  Using
Eqs.~(\ref{27}) and (\ref{28}), we find
\begin{eqnarray}
&&\partial  \ln[\sigma(\alpha,{s}_{\scriptscriptstyle 0},{\Phi}_{\scriptscriptstyle 0})] /\partial \alpha \cong
\frac{1}{2}\kappa({s}_{\scriptscriptstyle 0}) L \nonumber \\ && \times \eta(\alpha,{s}_{\scriptscriptstyle
0},{\Phi}_{\scriptscriptstyle 0}) \ln[1/{\eta(\alpha,{s}_{\scriptscriptstyle 0},{\Phi}_{\scriptscriptstyle 0})} ] \sin(\alpha) ,
\label{29}
\end{eqnarray}
to leading order.  In other words, the fractional azimuthal variations of charge density on the equipotential surface vanish as
$\eta(\alpha,{s}_{\scriptscriptstyle 0},{\Phi}_{\scriptscriptstyle 0}) \ln[\eta(\alpha,{s}_{\scriptscriptstyle
0},{\Phi}_{\scriptscriptstyle 0})]$.  Not surprisingly, this is the same rate at which the fractional variations of the diameter
vanish [cf. Eq.~(\ref{27})], and significantly more rapid than the rate at which the linear charge density of the filament tends
to uniformity.

The above results allow us to conclude that as the diameter of the filament decreases, the azimuthal nonuniformities in diameter
and surface charge density disappear relatively rapidly, confirming our earlier statement to that effect.  Equivalently, we can
conclude that in the limit of vanishing diameter, the azimuthal component of the electric field on the surface of a thin, circular cylinder vanishes as required.

Let us summarize the above results:

\textit{The density of charge along a regular conducting filament of fixed diameter tends to uniformity according to
Eq.~(\ref{24}), except near endpoints, as the ratio of its diameter to the smaller of its length and minimum radius of curvature
tends to zero.}

This is the main result of this paper, and as stated earlier, generalizes the result of Maxwell and others for straight
conductors to curved filaments. As such, it confirms the intuition that the charge distribution along a sufficiently smooth,
charged filament basically follows that of a straight one, and similarly converges to uniformity as the diameter of the filament
approaches zero. This is so despite the fact that there's an azimuthal variation of charge density on the surface of the
conductor at points of non-zero curvature which serves to balance the azimuthal component of the electric field generated by
deviation from axial symmetry.

\subsection{Charge Distribution Revisited Energetically}
The tendency of charges in equilibrium on thin, conducting conductors to become uniformly distributed is thus seen to be robust
and survive the generalization to curved filaments.  We arrived at this result in the preceding section by a detailed
consideration of the electric field components and equilibrium of charges, together with the boundary condition that all points
of a conducting body must be at the same potential.

At this juncture it is instructive to reconsider the distribution of charge along the filament \textit{ab initio} using energy
methods.  Specifically, we will look for that distribution of charge on the filament which minimizes the electrostatic energy,
given that charges are free to move within a conducting medium.

As a preliminary step, we will provide a proof that the state of minimum energy for an assembly of charged conductors is one of
uniform electrostatic potential throughout each conducting body \cite{equip}.  Consider a number of charged, insulated
conductors in electrostatic equilibrium \cite{grnded}.  The charge on the $i$th conductor will be denoted by ${Q}_{i}$ and the
region it occupies by ${\mathcal{R}}_{i}$.  Since the charges are confined to the conductors, we may express the electrostatic
energy of the system as
\begin{equation}
W={\sum}_{i}\,\, \frac{1}{2} \,\,{\int}_{{\mathcal{R}}_{i}} {d}^{3}r \rho(\mathbf{r})\Phi(\mathbf{r}),  \label{30}
\end{equation}
where $\rho(\mathbf{r})$ is the (volume) charge density and the integral is over the indicated region. Furthermore, since the
total charge on each conductor is given and fixed, we have
\begin{equation}
{\int}_{{\mathcal{R}}_{i}} {d}^{3}r \rho(\mathbf{r})={Q}_{i}.  \label{31}
\end{equation}

Rewriting Eq.~(\ref{30}) by expressing $\Phi$ in terms of $\rho$, we find
\begin{equation}
W={\sum}_{i}\,\, \frac{1}{2}k \,\,{\int}_{{\mathcal{R}}_{i}} {d}^{3}r {d}^{3}r'
\rho(\mathbf{r}){|\mathbf{r}-\mathbf{r'}|}^{-1}\rho(\mathbf{r'}).  \label{32}
\end{equation}
The problem before us then is to minimize $W$ in Eq.~(\ref{32}) subject to the constraints expressed in Eq.~(\ref{31}).
Following standard procedure, we associate a Lagrange multiplier ${\mu}_{i}$ with the $i$th constraint and seek to minimize the
quantity
\begin{eqnarray}
{\sum}_{i}\,\, \frac{1}{2}k \,\,{\int}_{{\mathcal{R}}_{i}} {d}^{3}r {d}^{3}r' \rho(\mathbf{r}){|\mathbf{r}-\mathbf{r'}|}^{-1}\rho(\mathbf{r'}) \nonumber \\
-{\sum}_{i}\,\,{\mu}_{i}[{\int}_{{\mathcal{R}}_{i}} {d}^{3}r \rho(\mathbf{r})-{Q}_{i}]  \label{33}
\end{eqnarray}
with respect to variations in the multipliers $\{ {\mu}_{i} \}$ as well as the charge density $\rho(\mathbf{r})$.

Setting the first-order variation with respect to $\rho(\mathbf{r})$ equal to zero gives
\begin{equation}
k{\int}_{{\mathcal{R}}_{i}} {d}^{3}r' \rho(\mathbf{r'}){|\mathbf{r}-\mathbf{r'}|}^{-1}=\Phi(\mathbf{r})= {\mu}_{i},\,\,\,
\mathbf{r}\in {\mathcal{R}}_{i}.  \label{34}
\end{equation}
The analogous equations for the multipliers simply reproduce the constraint equations of Eq.~(\ref{31}). Note that if
$\mathbf{r}\notin {\mathcal{R}}_{i}$ for any $i$, we get $0=0$ instead of Eq.~(\ref{34}).

We conclude from Eq.~(\ref{34}) that electrostatic energy is least when each ${\mathcal{R}}_{i}$ is an equipotential region,
i.e., there's a uniform potential throughout the body of each conductor.  This immediately implies the vanishing of the electric
field as well as the charge density at all but surface points of each conductor.  At surface points, on the other hand, the
equipotential condition implies that the electric field is normal to the surface of the conductor. We have thus arrived at the
familiar results of the electrostatics of conductors by energy minimization \cite{minimum}.

At this point the argument continues as in \S IIIB and culminates in Eq.~(\ref{24}), which conveys the inverse-logarithmic
approach of charge density to uniformity.   We have thus recovered the main result of this paper by means of energy
considerations.
\section{Examples \& Numerical Results}
To illustrate the main results of the foregoing analysis, we shall present the cases of straight and circular ideal filaments as
examples of Eqs.~(\ref{19}), (\ref{20}), and (\ref{23}) in the following subsection, and a detailed numerical and graphical
study of an elliptical conducting filament demonstrating the approach of charge density to uniformity with vanishing diameter,
Eq.~(\ref{24}), in \S IVB.
\subsection{Straight and Circular Filaments}
Our objective here is an analytical calculation of the fields of ideal straight and circular filaments for specified charge
densities, and a comparison of their asymptotic behavior in the limit of $\epsilon \rightarrow 0$ with those of Eqs.~(\ref{19})
and (\ref{20}).  As in the derivation of these equations, we will present the limiting behavior as an asymptotic, singular part that depends on
$\epsilon$, plus a remainder part that is finite and independent of $\epsilon$.  we will also verify the asymptotic formula for
capacitance, Eq.~(\ref{23}).

We shall start with the simplest example of a charge filament, namely a straight line of uniform charge density $\lambda$ extending from $z=-L/2$ to $z=+L/2$.  Then a straightforward calculation gives \cite{lit}
\begin{equation}
\Phi(z,\epsilon)=k \lambda \ln \Biggl \{ \frac{+(\frac{1}{2}-\zeta)+[{(\frac{1}{2}-\zeta)}^{2} +{\epsilon}^2]^{1/2}}
{-(\frac{1}{2}+ \zeta)+{[{(\frac{1}{2}+\zeta)}^{2} +{\epsilon}^{2}]}^{1/2}} \Biggr\}, \label{35}
\end{equation}
where $\zeta=z/L$.  This result can be used to find the asymptotic behavior of the potential as ${\epsilon \rightarrow 0}$:
\begin{equation}
\Phi(z)=k \lambda [\ln(1-4z^2/L^2)+2\,{\lim}_{\epsilon \rightarrow 0} \ln({\epsilon }^{-1})], \label{36}
\end{equation}
for the electrostatic potential at point $z$ of the filament,  and
\begin{eqnarray}
\mathbf{E}(\alpha,z)=8k \lambda z{(L^2-4z^2)}^{-1}{\boldsymbol{\hat{\tau}}}  +{\lim}_{\epsilon \rightarrow 0} 2k
\lambda{({\epsilon L})}^{-1} {\mathbf{\hat{n}}}(\alpha), \nonumber \\  \label{37}
\end{eqnarray}
for the electric field, valid for points not too close to the ends of the filament \cite{notclose}.  One can verify by
inspection that these results agree with their general counterparts, Eqs.~(\ref{20}) and (\ref{19}), as well as (\ref{23}), for
the case of a uniformly charged straight filament.

The second example is also a straight filament but with a uniform dipole density, i.e., $\lambda(z)=g z$, where $g$ is a
constant.  Then we find by straightforward integration
\begin{eqnarray}
&&\Phi(z,\epsilon)=kg z \ln \Biggl \{ \frac{+(\frac{1}{2}-\zeta)+[{(\frac{1}{2}-\zeta)}^{2} +{\epsilon}^2]^{1/2}} {-(\frac{1}{2}+ \zeta)+{[{(\frac{1}{2}+\zeta)}^{2} +{\epsilon}^{2}]}^{1/2}} \Biggr\} \nonumber \\
&& +k g L  \Bigl\{ \bigl[{(\frac{1}{2}-\zeta)}^{2} +{\epsilon}^2\bigr]^{1/2} - \bigl[{(\frac{1}{2}+\zeta)}^{2}
+{\epsilon}^2\bigr]^{1/2} \Bigr\}. \,\,\,\,\,\, \label{38}
\end{eqnarray}
Taking the limit of this equation as ${\epsilon \rightarrow 0}$, we find
\begin{equation}
\Phi(z)=kg \bigl[ \ln(1-4z^2/L^2)-2+2\,{\lim}_{\epsilon \rightarrow 0} \ln({\epsilon }^{-1})\bigr]z, \label{39}
\end{equation}
for the electrostatic potential at point $z$ of the filament,  and
\begin{eqnarray}
&&\mathbf{E}(\alpha,z)=k g [8z^2/(L^2-4z^2)-\ln(1-4z^2/L^2)+2]{\boldsymbol{\hat{\tau}}} \nonumber \\ &&+2k g\,{\lim}_{\epsilon
\rightarrow 0}[ \ln({\epsilon }){\boldsymbol{\hat{\tau}}} +z{({\epsilon L})}^{-1} {\mathbf{\hat{n}}}(\alpha)], \label{40}
\end{eqnarray}
for the electric field, valid for points not too close to the ends of the filament.  As in the above, inspection shows that
these results agree with the general ones given in Eqs.~(\ref{20}) and (\ref{19}).

Our third example is the case of a circular filament of radius $R$ and uniform charge density ${\lambda}$, centered at the
origin of a polar coordinate system $(r,\theta)$.  By symmetry, the electrostatic potential of such a filament is independent of
$s=R\theta$ and its axial electric field (i.e., the component parallel to $\boldsymbol{\hat{\tau}}$; see Fig. 1) is identically
zero.  On the other hand, unlike straight filaments, the potential of a circular one has an azimuthal variation and a
corresponding electric field component, as will be seen below.

Using the geometry depicted in Fig. 1, we find for the electrostatic potential of a circular filament
\begin{eqnarray}
&&\Phi(\epsilon, \alpha)=4k{\lambda}{ \{ 4[1-\epsilon \cos(\alpha)] + {\epsilon}^2 \} }^{-1/2} \nonumber \\
&&\times K \big[ { \big( [4(1-\epsilon \cos(\alpha)] / \{ 4[1-\epsilon \cos(\alpha)] + {\epsilon}^2 \} \big) } \big],
\,\,\,\,\,\,\,  \label{41}
\end{eqnarray}
where $K(\cdot)$ is the complete elliptic integral of the first kind defined as
\begin{equation}
K(u)={\int}_{0}^{\pi/2}d\theta {[1-u \,{\sin}^2(\theta)]}^{-1/2}. \label{42}
\end{equation}
Note that the azimuthal angle $\alpha$ equals zero at points $r=R-\epsilon,\,\,0 \leq \theta \leq 2 \pi$, and it equals $\pi$ at points  $r=R+\epsilon,\,\,0 \leq \theta \leq 2 \pi$.

The limiting behavior of the potential in Eq.~(\ref{41}) can be found using the asymptotic formula $K(u) \rightarrow
\ln[{4/(1-u)}^{1/2}]$, $u \rightarrow 1$.  Keeping terms up to order $\epsilon \ln(\epsilon)$, we find
\begin{equation}
\Phi(\alpha)\cong 2k \lambda \bigl\{[1+\frac{1}{2}\cos(\alpha)\epsilon] \ln(8/\epsilon) -\frac{1}{2}\cos(\alpha)\epsilon
\bigr\}.  \label{43}
\end{equation}
Up to terms of order $\ln(\epsilon)$, the right-hand side of Eq.~(\ref{41}) reduces to $2k \lambda \ln(8/\epsilon)$, in
agreement with Eq.~(\ref{20}).  Note also that the capacitance of the circular filament is given by $2 \pi R \lambda /\Phi$,
which leads to $c(\epsilon) \rightarrow [2k  \ln(8/\epsilon)]^{-1}$ for its capacitance per unit length, in agreement with Eq.~(\ref{23}) as $\epsilon \rightarrow 0$.

The electric field of the circular filament is found by calculating the negative gradient of Eq.~(\ref{43}). The result is
\begin{eqnarray}
&& \mathbf{E}(\alpha,\theta)=(k\lambda/R)\Bigl\{ \bigl[ \cos(\alpha)[2-\ln(8)] \nonumber \\ &&+ {\lim}_{\epsilon \rightarrow 0}[\cos(\alpha)\ln(\epsilon) +2{\epsilon}^{-1}] \bigr] {\mathbf{\hat{n}}}(\alpha,\theta) \nonumber \\
&& -\sin(\alpha)[1-\ln(8)+ {\lim}_{\epsilon \rightarrow 0} \ln(\epsilon)] \boldsymbol{\hat{\alpha}}(\alpha,\theta) \Bigr\},
 \label{44}
\end{eqnarray}
which, given the substitutions $L \rightarrow R$ and $\kappa \rightarrow 1/R$, agrees with Eq.~(\ref{19}).

It is instructive to find the azimuthal variations of surface charge density for the circular filament using Eq.~(\ref{44}).
Following the procedure given in \S IIIB, we find
\begin{equation}
\partial \ln[\sigma(\alpha,\theta)] /\partial \alpha ={\lim}_{\epsilon \rightarrow 0}\frac{1}{2} \sin(\alpha) \epsilon \ln({\epsilon}^{-1}),  \label{45}
\end{equation}
in agreement with the general result in Eq.~(\ref{29}).  In particular, this  result confirms our conclusions in \S IIIB that
azimuthal nonuniformities vanish relatively rapidly as $\epsilon \rightarrow 0$.

As stated earlier, the three cases treated above serve to exemplify the general asymptotic results derived in \S II and III.
\subsection{Elliptical Filament}
To illustrate the approach of charge density to uniformity for a curved filament whose axial field (the electric field component
parallel to $\boldsymbol{\hat{\tau}}$; see Fig. 1) does not vanish identically due to symmetry, we will consider an elliptical
filament and numerically investigate the behavior of the charge density and electrostatic potential as $\epsilon \rightarrow 0$.
We will also calculate the capacitance of the filament as a function of its diameter (thickness) in that limit.

Consider an the elliptical filament of axes $(a,b),\,\,a=2b,$ centered at the origin of a polar coordinate system $(r,\theta)$
with its major axis at $\theta=0\,\, \& \,\, \pi$.  Using the coordinate system depicted in Fig. 1, we will numerically
calculate the axial component of the electric field for the first quadrant, $0 \leq \theta \leq \pi/2$, at the azimuthal angle
$\alpha = \pi/2$, with $\epsilon$ decreasing from ${10}^{-4}$ to ${10}^{-12}$. Recall that the filament thickness is given by
$2L\epsilon$, where in this case $L=(b^2/a)$ is the minimum radius of curvature for the ellipse which obtains at points
$\theta=0\,\, \& \,\, \pi$, the points of extremal distance from the foci.  Note that $\epsilon={10}^{-12}$ corresponds to subatomic dimensions for a filament of, say, $a=1$ m.

Starting with the general representation of the axial field in Eq.~(\ref{5}), we make the replacements $\lambda(s)ds \rightarrow
\lambda(\theta')d\theta'$, $\mathbf{R}({s}_{\scriptscriptstyle 0}) \rightarrow
\mathbf{R}(\theta)=a\cos(\theta)\hat{x}+b\sin(\theta)\hat{y}+\epsilon (b^2/a)\hat{z}$, $\mathbf{R}({s}) \rightarrow
\mathbf{R}(\theta')=a\cos(\theta')\hat{x}+b\sin(\theta')\hat{y}$, where we have associated Cartesian unit vectors with the polar
plane in the standard manner.  Note that $\lambda(\theta')=\lambda(s)\,ds/d\theta'$, where $ds/d\theta'=[a^2
{\sin}^{2}(\theta')+b^2 {\cos}^{2}(\theta')]^{1/2}$.

The three unit vectors constituting the triad of Fig. 1 correspond here to $\boldsymbol{\hat{\tau}}(\theta)$ as the unit tangent to the ellipse
in the counter-clockwise direction, $\boldsymbol{\hat{\nu}}(\theta)$ as the inward unit normal to the ellipse in the polar
plane, and $\boldsymbol{\hat{\beta}}=\hat{z}$ as the unit binormal, which is a fixed vector in this case.  Note that we have chosen to
look at the potential and the $\tau$-component of the electric field at the azimuthal angle $\alpha=\pi/2$, corresponding to the
filament's ``ridge line.''  However, the choice of this angle makes little difference for a sufficiently thin filament.

For the charge density $\lambda(s)$, we assume the form ${\lambda}_{\scriptscriptstyle 0}+\Delta {\lambda}(s)/\ln(\epsilon)$
suggested by Eq.~(\ref{24}), and approximate $\Delta {\lambda}(s)$ by a three-term Fourier series. Ideally, the charge density would be determined by the requirement that the potential $\Phi(\theta, \alpha, \epsilon)$ is uniform, in which case we would find an $\epsilon$-dependent result, to wit, $\Delta {\lambda}(s,\epsilon)$.  We have determined the three Fourier coefficients by minimizing the variations of the potential at $\epsilon=10^{-12}$, and expect the potential to be reasonably uniform at $\epsilon=10^{-12}$, and less so at other values of $\epsilon$.

The primary objective of the numerical investigation is to show that as the diameter of the filament decreases, the charge density tends to uniformity while the potential remains essentially uniform.  The results shown here  confirm this expectation as the relative variations of the charge density and potential are found to be of the order of $10^{-2}$ and $10^{-4}$ for $\epsilon=10^{-12}$.  It is worth recalling here that the approach of charge density to uniformity is inverse-logarithmic, hence extremely slow.

\begin{figure}
\includegraphics[]{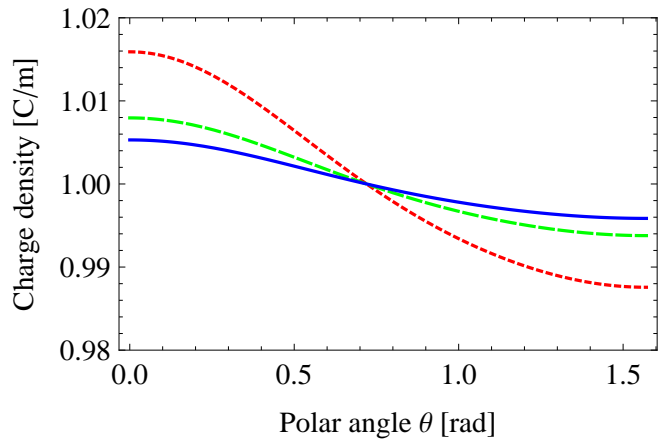}
\caption{Approach of the linear charge density to uniformity for an elliptical filament.  The three graphs show the charge density for the first quadrant of the elliptical filament, at  $\epsilon=10^{-4}$ (short dash, red), $\epsilon=10^{-8}$ (long dash, green), and $\epsilon=10^{-12}$ (solid, blue).  Note the logarithmically slow rate of convergence to uniformity.}
\label{fig2}
\end{figure}

Figure 2 is a plot of charge density at three values of $\epsilon$, namely $10^{-4}$ (short dash), $10^{-8}$ (long dash), and $10^{-12}$ (solid), for which the relative deviations of charge density from uniformity are expected to be in the ratio of $1/4:1/8:1/12$ respectively.  The slow rate of convergence to uniformity well known from studies of straight filaments is clearly in evidence here.

\begin{figure}
\includegraphics[]{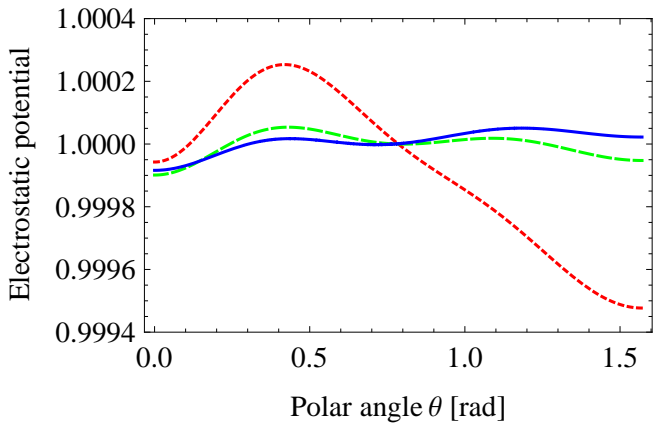}
\caption{Approach of the electrostatic potential to uniformity for an elliptical filament.  The three graphs show the
electrostatic potential, normalized to unity at $\theta=\pi/4$, for the first quadrant of the elliptical filament, at
$\epsilon=10^{-4}$ (short dash, red), $\epsilon=10^{-8}$ (long dash, green), and $\epsilon=10^{-12}$ (solid, blue).}
\label{fig3}
\end{figure}

The behavior of the electrostatic potential for the same values of $\epsilon$  as above are shown in Fig. 3, where we have normalized the potential to unity at $\theta=\pi/4$.  This is of course the quantity that must be uniform for a conducting filament.  Indeed the potential varies about $0.01\%$ at $\epsilon=10^{-12}$ and $0.1\%$ at $\epsilon=10^{-4}$.  Given the approximate nature of our fit to $\Delta {\lambda}(s,\epsilon)$, we find the results shown in Fig. 3 in agreement with our expectations.

It is worth recalling that a strictly uniform charge density will in general produce a nonuniform potential along the filament regardless of how small its diameter is.  This nonuniformity is reflected in the axial part of $\mathbf{\tilde{E}}(\alpha,{s}_{\scriptscriptstyle 0})$ in Eq.~(\ref{19}) for the general case, and exemplified by the first term of Eq.~(\ref{37}) for the straight filament, both of which represent the axial electric field in the limit of $\epsilon =0$ for the case of uniform charge density.   That such a residual axial electric field persists in the limit of vanishing diameter is of course the puzzle raised in Ref. \cite{djgyli} and explained in detail in Ref. \cite{jdj1}.  Whether straight or curved, the residual axial field is caused by asymmetric contributions from charges not in the immediate neighborhood of a given point and requires an appropriate amount of local dipole density, $\Delta \lambda(s,\epsilon)$, to counteract it.  Points that receive symmetric contributions from all charges, such as the midpoint of a straight filament and points $\theta=0$ and $\theta=\pi/2$ of the elliptical filament (as well as at the equivalent points $\theta=\pi$ and $\theta=3\pi/2$ not shown in the figures), are therefore exceptions to the general statement.  As seen in Fig. 3, the potential is stationary at these exceptional points for the three values of $\epsilon$ represented.

\begin{figure}
\includegraphics[]{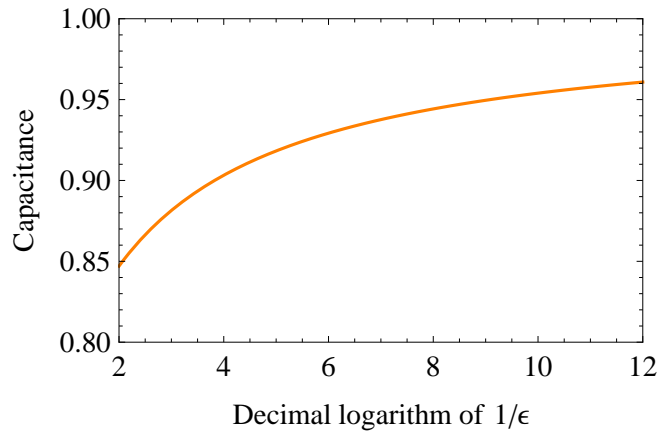}
\caption{Capacitance per unit length of an elliptical filament relative to a circular one as its diameter shrinks by $10$ orders of magnitude; see text.}
\label{fig4}
\end{figure}

Finally, remembering that Maxwell's original contribution focused on the capacitance of the straight filament, we show the capacitance per unit length of the elliptical filament (normalized to that of a circular one whose radius matches the minimum radius of curvature of the elliptical filament) in Fig. 4 for $\epsilon$ ranging from $10^{-2}$ to $10^{-12}$.  Recall from \S IIC, Eq.~(\ref{23}), that as $\epsilon \rightarrow 0$, the capacitance per unit length of a regular filament approaches a universal form that only depends on its diameter.  This result is born out by the behavior seen in Fig. 4, which also echoes the slow rate of convergence to uniformity characteristic of the problem.

We have posted an interactive version of the computer program used for the above calculations for use by interested readers \cite{code}.  It is a code written in \textit{Mathematica} and annotated by explanatory comments.  The user can manipulate the results or change the parameters, subject to the limitations resulting from the nearly singular nature of the integrations.

\section{Concluding Remarks}
The exterior field of an infinitely long, conducting, charged, circular cylinder does not depend on its diameter, and is therefore the
same as if its charges were to be concentrated on its axis.   When finite length effects for a thin conductor are taken into account,
the correction terms to the uniform charge density of the infinite case turn out small, except near the endpoints, and vanish as
the inverse of the logarithm of the ratio of the diameter to length of the conductor \cite{jdj1,jdj2}.  What we have found here
is that the same conclusion holds for the charge density along a sufficiently smooth conducting filament even if curved, albeit
with appropriate modifications arising from the nonzero curvature of the filament.  The intuition behind this result is the
observation that at a point where the diameter of the filament is sufficiently small compared to either its length or radius of
curvature, the filament is locally indistinguishable from a straight one. In both curved and straight cases, the leading
behavior arises from local charge and dipole densities while the correction terms originate from charges farther away.

Beyond the main uniformity result, the two cases differ in important details.  For example, while the charge distribution on any straight
filament can be described by an appropriate scaling of a universal function, such is not the case with curved filaments which
can assume an enormous variety of forms.  In particular, a curved filament can be closed, in which case complications introduced
by endpoints would not be present.

The regularity conditions imposed on the geometry of the filament in \S IIA are not mere mathematical niceties since the uniformity of charge distribution on the filament will break down in the presence of a corner or a kink on the filament \cite{corner}.  However, although we have excluded such points altogether, it should be noted that the uniformity results of this paper are still valid at any point of the filament whose neighborhood conforms to the smoothness conditions, even if they are violated
elsewhere.

{}


\begin{thebibliography}{}
\bibitem{djgyli} D. J. Griffiths and Y. Li, ``Charge density on a conducting needle,'' Am. J. Phys. \textbf{64}, 706 (1996).
\bibitem{good} R. H. Good, ``Comment on `Charge density on a conducting needle,' '' Am. J. Phys. \textbf{65}, 155 (1997).
\bibitem{and} M. Andrews, ``Equilibrium charge density on a conducting needle,'' Am. J. Phys. \textbf{65}, 846 (1997).
\bibitem{jdj1} J. D. Jackson, ``Charge density on thin straight wire, revisited,'' Am. J. Phys. \textbf{68}, 789 (2000).
\bibitem{jdj2} J. D. Jackson, ``Charge density on a thin straight wire: The first visit,'' Am. J. Phys. \textbf{70}, 409 (2002).
\bibitem{jcm} J. C. Maxwell, ``On the electrical capacity of a long narrow cylinder, and of a disk of sensible thickness,'' Proc. London Math. Soc. \textbf{IX}, 94 (1878).
\bibitem{smooth} Note that this regularity condition (existence of a non-vanishing, continuously differentiable tangent) is stronger than
smoothness, which requires the existence of a non-vanishing, continuous tangent.
\bibitem{tbp}  Stated in a mathematically proper manner, this is our \textit{definition} of a curved, circular cylinder.
\bibitem{regular} We will refer to a charged filament statisfying these requirements as simply ``regular.''  Needless to say, smoothness characteristics of $\lambda(s)$ under equilibrium conditions are determined by those of $\mathbf{R}(s)$ and are not
an independent issue.
\bibitem{lcdev}  The device of using a line charge to describe the fields of the physical filament was also used by Andrews \cite{and} and Jackson \cite{jdj1}.
\bibitem{lcd} Here the length parameter for the physical conductor is defined to be the same as that of the curve $\mathbf{R}(s)$ within.
\bibitem{rowley} R. J. Rowley, ``Finite line of charge,'' Am. J. Phys. \textbf{74}, 1120 (2006).
\bibitem{wrs} W. R. Smythe, \textit{Static and Dynamic Electricity} (McGraw-Hill, New York, 1968), 3rd ed., pp. 123-124.
\bibitem{recall}  Recall that $2\eta(\alpha,{s}_{\scriptscriptstyle 0},{\Phi}_{\scriptscriptstyle 0})$ is the diameter of the filament in units of $L$, the smaller of the length and minimum radius of curvature of the filament, and that $\kappa({s}_{\scriptscriptstyle 0})L$ is a pure number that never exceeds unity.
\bibitem{equip}  This result is traditionally established by the argument that electrostatic equilibrium precludes the motion of charges within a conductor.
\bibitem{grnded} If present, grounded conductors can be regarded as one with the Earth and treated as another insulated conductor for the purpose of this derivation.
\bibitem{minimum}  The vanishing of the first-order variations guarantees a stationary configuration, not necessarily a minimum.  However, the minimum nature of the result here is clear on physical grounds, and can also be ascertained mathematically by examining the second-order variations.
\bibitem{lit} Calculations of the fields of straight filaments appear in various forms in the literature.  See, e.g., Refs. \cite{and, jdj1}.
\bibitem{notclose} Specifically, we require that $1/2-|z/L| \gg \epsilon$.
\bibitem{corner} Mathematically, the controlling scale parameter $L$, the
smaller of the length and minimum radius of curvature of the filament, would cease to exist, or perhaps vanish, at such points,
in violation of our regularity conditions in \S IIA.    Physically, a ``corner'' essentially amounts to a point where the radius
of curvature of the filament is smaller than, or comparable to, its diameter, so that $\epsilon\gtrsim 1$, far from the limit
required for uniformity.
\bibitem{code} The \textit{Mathematica 7} program is posted to the URL \url{http://www.csus.edu/indiv/p/partovimh/pg.nb}.

\end{thebibliography}
\end{document}